# THE EFFECTS OF IMPROPER LIGHTING ON PROFESSIONAL ASTRONOMICAL OBSERVATIONS

Ferdinando Patat[1,2]

[1]European Organization for Astronomical Research in the Southern Hemisphere, Germany
[2]International Astronomical Union

**ABSTRACT**

Europe and a number of countries in the world are investing significant amounts of public money to operate and maintain large, ground-based astronomical facilities. Even larger projects are under development to observe the faintest and most remote astrophysical sources in the universe. As of today, on the planet there are very few sites that satisfy all the demanding criteria for such sensitive and expensive equipment, including a low level of light pollution. Because of the uncontrolled growth of incorrect illumination, even these protected and usually remote sites are at risk. Although the reasons for intelligent lighting reside in energy saving and environmental effects, the impact on scientific research cannot be neglected or underestimated, because of its high cultural value for the progress of the whole mankind. After setting the stage, in this paper I review the effects of improper lighting on professional optical and near-UV astronomical data, and discuss the possible solutions to both preserve the night sky natural darkness and produce an efficient and cost-effective illumination.

## 1. INTRODUCTION

It is often said that astronomers hate light. Well, this is not true. They actually love light (and more in general electromagnetic radiation), because this is the only way they have to perceive and study the universe. It is light, with its extreme speed, that can fill the unconceivable gap that separates the astronomer from her object of study. Light that in some cases has to travel billions of light years before reaching our telescopes.
In the last twenty years, ground-based astronomical facilities, placed on the most remote corners of our planet, have been growing significantly. Furthermore, large efforts are currently being undertaken for the construction of 30 to 40 meter telescopes, to start operations before 2020. These billion-euro projects (**E**uropean **E**xtremely **L**arge **T**elescope, **G**iant **M**agellan **T**elescope, **T**hirty **M**eter **T**elescope) aim at addressing a number of fundamental questions in modern astrophysics and cosmology. And these include the question of questions: are there other life forms in the universe?

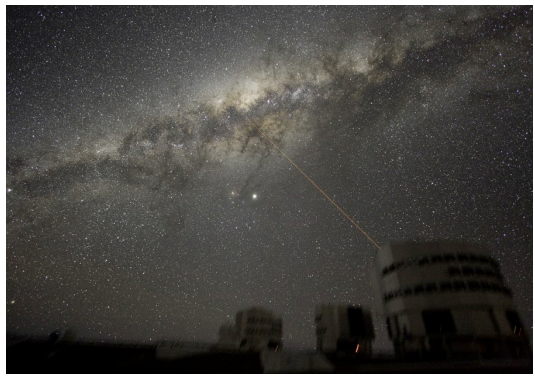

*Figure 1.* The Milky Way above the ESO observatory on Cerro Paranal (Atacama desert, Chile. Photo credit: ESO). A natural show concealed to most European citizens.

One might be tempted to think all this research activity is the privileged domain of a limited, elite group of individuals. Indeed, here I am writing on behalf of *just* ten thousand professional astronomers that are affiliated to the International Astronomical Union. However, this is only a reductive view of a much wider picture. Think of the millions of amateur astronomers all over the world. Consider the deep roots of Astronomy in history and culture, the effects astronomical findings had and have on our way of perceiving and conceiving the world. Remember the consequences of the findings of Galileo, Copernicus, and Einstein. Imagine what the discovery of an extra-solar planet similar to our Earth will mean. Try to figure what the detection of life will entail for us all…

## 2. THE NATURAL NIGHT SKY

Even at dark sites, far from any artificial light contamination, the night sky is not completely dark. In fact, when one is observing from the ground, there are several sources that contribute to its brightness, some of which are of extra-terrestrial nature (e.g. unresolved stars/galaxies, diffuse galactic background, zodiacal light), and others are due to atmospheric phenomena (airglow and auroral activity in the upper Earth's atmosphere) [1]. While the extra-terrestrial components vary only with the position on the sky, and are therefore predictable, the terrestrial ones are known to depend on a large number of parameters (season, geographical position, solar cycle and so on), which interact in a largely unpredictable way. In fact, airglow contributes to a significant fraction of the global, optical night sky emission (up to ~50%), and hence its variations have a strong effect on the overall brightness, which changes from wavelength to wavelength.

In the blue, the spectrum is rather featureless and it is characterized by the so-called airglow pseudo-continuum, which arises in layers at a height of about 90-100 km (mesopause), and extends all the way from 400nm to 700nm. All visible emission features, which become particularly marked below 400nm and largely dominate the UV, are due to Herzberg and Chamberlain bands of molecular oxygen ($O_2$) [1].

The visible is chiefly dominated by oxygen 558nm and to a lesser extent by sodium D doublet 590nm and OI 630,636 nm doublet. Besides the aforementioned pseudo-continuum, several oxydryl (OH) Meinel vibration-rotation bands are also present in this spectral window. All these features are known to be strongly variable and show independent behavior. In fact, OI 558nm, which is generally the brightest emission line in the optical sky spectrum, arises in layers at an altitude of 90 km, while OI 630,636 nm is produced at 250-300 km. The OH bands are emitted by a layer at about 85 km, while the Na ID is generated at about 92 km, in the so called sodium-layer which is used by laser guide star adaptive optic systems. In particular, OI 630,636 shows a marked and complex dependency on geomagnetic latitude, which turns into different typical line intensities at different observatories, and it is known to undergo abrupt intensity changes on the time scale of hours.

In the red, strong OH Meinel bands begin to appear, while the pseudo-continuum remains constant. Finally, as one approaches the near-IR, the spectrum is dominated by OH Meinel bands. Due to these phenomena, the sky is much darker in the blue-visible, and it becomes progressively brighter as one goes to the red. Then, when one enters the infra-red domain, the thermal background becomes more and more the dominant source of radiation, making ground-based observations very difficult.

The dark time sky brightness shows strong variations within the same night on the time scales of tens of minutes to hours. This variation is commonly attributed to airglow fluctuations. Moreover, as first pointed out by Lord Rayleigh, the blue and visible night sky brightness depends on solar activity. During a full sunspot cycle, it changes by about 60%, with the highest fluxes reached during the maximum of solar activity.

Besides the atmospheric airglow, there are other natural sources. Among these, the dominant one is the sun light diffused by dust distributed on the ecliptic plane, known as the zodiacal light. This can contribute up to 50% of the total brightness if one is observing at low ecliptic latitudes and, in general, it depends on the position of the sky where one is pointing. In dark sites, this is clearly visible also to the naked eye, as a diffused and wide cone, just after and before the evening and morning twilights respectively.

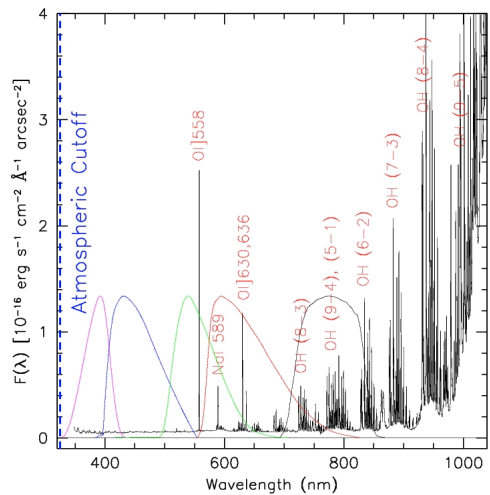

*Figure 2. Night sky spectrum obtained at Cerro Paranal (Chile) with the Focal Reducer low dispersion Spectrograph (FORS1) mounted at the 8.2m Kueyen VLT telescope [2]. Main emission line identifications are given. The dashed line on the left indicates the atmospheric cutoff, while the colored curves trace the U, B, V, R, and I astronomical passbands.*

The spectral range an optical, ground-based astronomer has access to is constrained by two natural phenomena: the atmospheric cutoff to the blue (which shields us from the venomous UV radiation emitted by the sun), and the enhanced natural sky brightness emerging in the near-IR. A quick look to Figure 2 immediately tells that the region between 320nm and 550nm is actually the darkest spectral interval accessible from the ground. This means that this is the region where the deepest images and spectra can be obtained. For this very reason, it deserves a special care.

For ground-based astronomy, the quality of the observing site is fundamental and this necessarily forces astronomers to place their telescopes in the most desolated and remote places of our planet. In fact, frontline optical and infrared observations require very good transparency conditions, very low humidity, a high number of clear nights, steady and non-turbulent winds and, of course, a dark sky. This has become even truer during the last fifty years, during which the growth of cities and industries, especially in Europe and North America, has added a new kind of environmental contamination, which is usually referred to as light pollution.

### 3. GOOD BYE, DARK SKY. THE DEVASTATING EFFECTS OF IMPROPER LIGHTING

The problem has become indeed serious: there is no dark spot in Europe, as the satellite night time images clearly show [3]. Practically almost no professional astronomical observations are carried out from continental Europe, and all large European telescopes have been placed either in the Canary Islands or in some remote places, usually in the northern and southern American continents. It is very sad to notice that the Milky Way is already a forgotten spectacle for many of us.

The effect of light pollution on astronomical observations is devastating [4].

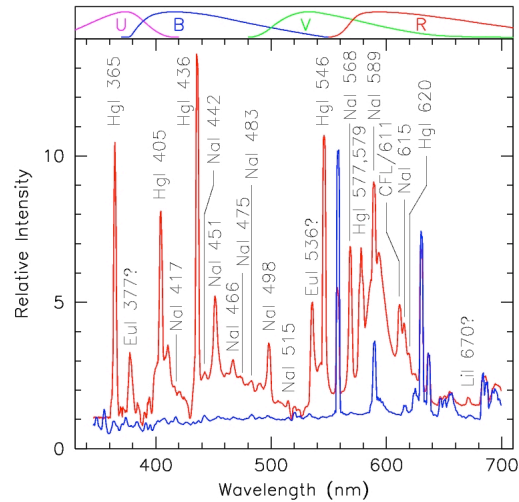

*Figure 3. Comparison between a night sky spectrum obtained in a light polluted site (red: Asiago Astrophysical Observatory, Italy) and a dark site (blue: ESO-Paranal, Chile [2]). Spectral line identifications for the main artificial features are given. The emissions generated by street lighting are clearly visible. They fall not only in the visible range (500-600nm), but also in the in blue and violet parts of the spectrum.*

Street lamps do not illuminate just the ground. Some radiation is reflected towards the sky, and in the most deprecated cases, even directly sent upwards by lamps with a bad design. When this light reaches the troposphere, it is reflected back to the ground and it overwhelms the natural night sky brightness, making the detection of faint and remote astronomical sources very difficult, if not impossible. What light designers often tend to ignore is that, due to its physical nature, the atmosphere produces what physicists call *forward scattering*. That is to say that when photons hit air molecules, they are deviated preferentially close to the incoming direction. This means that artificial light sent to the sky at low angles above horizontal can travel up to many hundreds of km before being finally scattered back to the ground [4].

Besides not being justified by any practical reason, sending light above horizontal is just a waste of energy, it contributes in a significant way to pollute the sky, and as a consequence it seriously hinders astronomical observations [4]. Currently, outdoors artificial illumination increases the sky luminosities chiefly through emission lines of mercury and sodium (which affect in particular the visual pass-band), but also through intense emissions in the blue and violet parts of the spectrum, clearly invading the scotopic domain. Other lines are produced by special elements (like rare earths), which are introduced in some sources (e.g. fluorescent lamps). The effects of artificial light contamination are clearly visible in Figure 3, where the night-time spectrum obtained in a light polluted European site (mount Ekar, Italy) is compared to the one obtained on Cerro Paranal (Atacama desert, Chile). The original scientific spectrum was obtained with the telescope pointing almost at zenith, in order to minimize the impact of atmospheric extinction. Notwithstanding this precaution, the sky spectrum shows overwhelming signs of artificial emission lines of mercury and sodium. The 1.8m Copernicus telescope is placed on the top of Mount Ekar, at about 1300 meters on the sea level. In spite of its relatively good location, the brightest feature in its night sky spectrum is the mercury line at 436nm, which is brighter than the natural oxygen line at 558nm, the most intense feature emitted by the night sky at these latitudes.

The presence of such lines causes the sky to become artificially brighter, and it turns into a disaster for broad-band imaging. In a relatively protected site like Mount Ekar, in the blue and visual pass-bands the enhancement is more than a factor of three. But when one goes close to a relatively big town, this degradation can reach a factor of twenty or more, causing many astronomical objects to be lost, not only for the visual observers, but also for the evolved amateurs equipped with modern digital cameras. A possible solution to the problem, at least for photographic and digital imaging, is the use of sky-suppression filters, which are designed to have a very low transmission in the spectral regions centered on the main mercury and sodium emission lines. For this reason, emission line street lighting (especially low- and high-pressure sodium) is definitely more astronomy-friendly than continuum sources, like LEDs (see also below) [5]. Of course, professionals and their extremely sensitive instruments are affected by light pollution too, which makes deep imaging a very difficult task, since the augmented sky background introduces an additional noise in the images, increasing the detection threshold. The contamination is a bit less of a problem for spectroscopy, since in that case one can still observe "in-between" the emission lines. Provided that street lamps have an emission line spectrum. With the advent of LEDs, however, this is not warranted anymore.

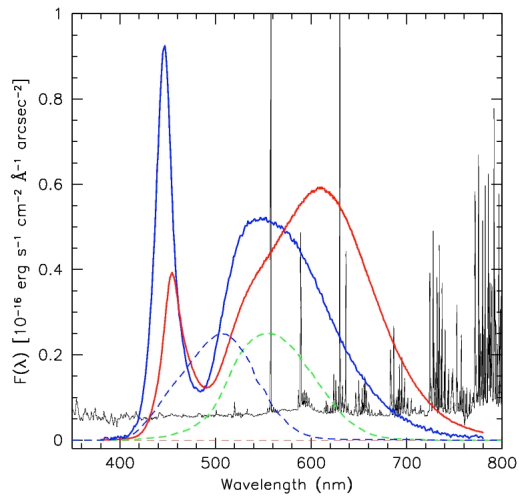

*Figure 4.* The natural night sky in the region 350-800 nm. The dashed curves trace the photopic (green) and scotopic (blue) passbands. The red and blue solid curves trace the $V(\lambda)$-weighted relative flux spectra of a white-warm (CCT=3000K; red), and a blue-white (CCT>5000K; blue) LED [6].

For these reasons, the deployment of cold-blue LEDs in street lighting (especially if not properly shielded), is going to cause a two-fold problem:

a) the LED blue-peak will invade the darkest portion of the natural night sky spectrum (see Figure 4);

b) at variance with line-emission sources the continuum produced by LEDs contaminates astronomical observations on a wide spectral range.

The amount of light emitted in the blue-peak is particularly significant for blue-white LEDs, with Correlated Colour Temperature (CCT) higher than 3000K. This peak falls in the scotopic region (see Figure 4), and therefore it is not really useful for night activities which require mesopic or even photopic illumination conditions.

Another very important aspect to be considered is the fact that the scattering efficiency within Earth's atmosphere obeys the Rayleigh law [4,5]. That is to say that the amount of upwards artificial lighting sent back to the ground is proportional to the fourth (!) power of the inverse of the wavelength. In simpler words, the atmosphere is significantly more efficient in reflecting back blue light than red light (this is also the reason why the daytime sky is so blue). The net effect of this physical fact is that blue sources are very efficient light polluters [4,5].

Cold-white LEDs pose a serious treat to astronomy.

## 4. CONCERNS, RECOMMENDATIONS AND CONCLUSIONS

The addition of any artificial light into the night-time environment has negative effects. As we have seen, the problems are primarily caused by sources emitting at wavelengths shorter than 500nm (violet and blue). Besides disrupting the natural behaviour of wildlife, and possibly degrading population's health (disturbance in melatonin production, disruption of circadian cycles, discomfort and increase of glare), it severely affects astronomy.

An argument often used to boost the deployment of LEDs for outdoors lighting is the implied energy savings and the corresponding decrease in green-house gas emission. This argument is valid, though, only under some assumptions. Mankind history is full of examples where new technology had allowed achieving the same result at a lower cost and lower energy consumption rate. However, this has very rarely turned into a global net saving. For instance, the amount of petrol and natural gas used for energy production has not decreased with time (the opposite is actually true), despite the undeniable technological progresses made in the field. It is arguable this will occur with the deployment of more energy-efficient light sources too.

If it is certainly true that with LEDs one can produce the same amount of light with much less energy, it is equally true that *with LEDs one can produce much more light with the same amount of energy*. What will happen is something in between these two extremes but, if anything, the global amount of light will certainly increase (and so will the amount of wasted light, unless proper actions are undertaken)[1]. During the CIE 2010 conference it was said many times that "*the future ahead of us is bright!*". Whilst this sentence was proclaimed with pride, it should rather raise serious worries about the attitude it betrays.

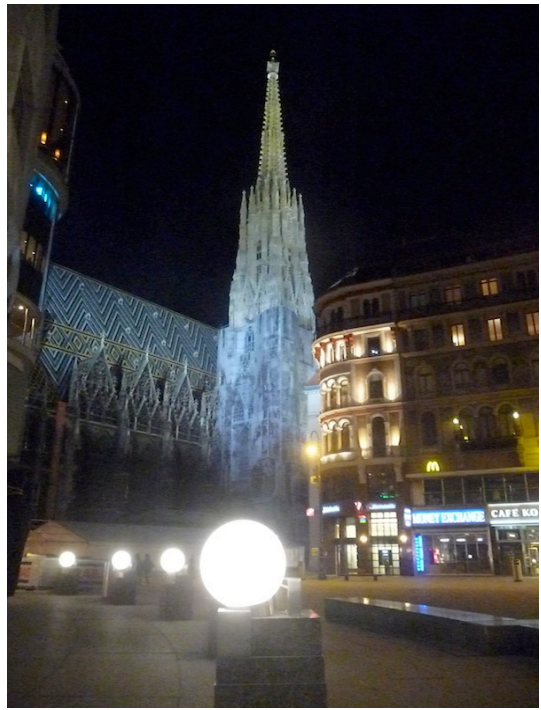

*Figure 5*. What is the purpose of these shining globes? Blind you? Hide St. Stephan's dome? Illuminate the heavens?

---

[1] I often refer to this as "My Grandmother's Theorem". When the CFL lamps appeared on the marked, and I told my Grandma that she could now save some money, she replied that as she was already accustomed to pay that bill, she would now rather have some lamps in the garden, and leave them on all night.

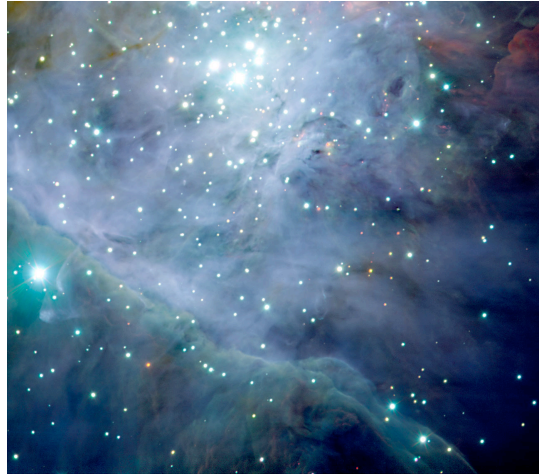

*Figure 6. Beauty meets science. A star forming region in the Orion Nebula (photo credit: ESO) [8].*

Because this is exactly what is going to happen, severely questioning the energy-saving argument. Visual and environmental effects must be understood, not just the potential for savings, especially if the underlying justification is [partially] flawed. Savings can be achieved also through a better lighting design (proper illumination levels and distribution), using modern remote control technology, removing unnecessary lights and deploying luminaries with better directional and glare control (see Figure 5).

These are the golden rules for an astronomy-friendly outdoors/street lighting:

a) Minimize the use of light sources with CCT>3000K. In general, limit emission below 500nm as much as possible.

b) Utilize control options (motion sensors, time dimming, including complete shutoff).

c) 0% **U**pward **L**ight **O**utput **R**atio (ULOR). That is to say, zero emission above horizontal!

d) Limit illumination to task specific areas and avoid over-illumination.

e) Use low-pressure sodium and (as a second choice) high-pressure sodium whenever possible.

In this respect, it might be useful to remember that high-pressure sodium is a very energy-efficient light source, still competitive with (if not better than) current technology LEDs [7].

The night sky must have impressed the human beings since the beginning of their evolution, at the very origin of mankind's culture. Its beauty continues to astonish everybody who turns her eyes to the heavens, no matter whether she is an astronomer or simply a lover of nature. Sadly, less and less people can nowadays enjoy this wonderful experience, which is unfortunately becoming a privilege.

The International Astronomical Union is confident that CIE will use its scientific authority to help us preserving the beauty of one of the most astounding shows nature ever gave us.

## ACKNOWLEDGEMENTS

I am grateful to the [courageous] organizers of CIE 2010 for inviting me to the represent the International Astronomical Union, and present this paper. I also wish to thank Andrej Mohar (Dark Sky Slovenia), Fabio Falchi (Cielo Buio, Italy), Elizabeth Alvarez del Castillo (National Optical Astronomy Observatory, USA), and Richard Wainscoat (University of Hawaii, USA) for their kind help and support during the preparation of this paper.

## AUTHOR

Dr. Ferdinando Patat
European Organization for Astronomical Research in the Southern Hemisphere
K. Schwarzschild-Str. 2, 85748 Garching b. München, Germany
fpatat@eso.org - www.eso.org/~fpatat